\documentclass[twocolumn,nofootinbib,preprintnumbers,superscriptaddress,amsmath,amssymb,PRL]{revtex4-1}
\usepackage{graphicx}
\usepackage{dcolumn}
\usepackage{float}
\usepackage{mathptmx, courier, pifont}
\usepackage[scaled=0.92]{helvet}
\usepackage[T1]{fontenc}
\usepackage{textcomp}
\usepackage{color}
\usepackage[normalem]{ulem}

\usepackage[dvipsnames]{xcolor}
\usepackage[colorlinks=true,urlcolor=Blue,linkcolor=Blue]{hyperref}
\usepackage[all]{hypcap}
\usepackage{makecell}

\usepackage{physics} 
\usepackage{multirow}

\begin{document}


\title{Precise determination of electron-capture $Q$ value of $^{113}$Sn decay related to electron neutrino mass measurements}
\author{Zhuang~Ge}\thanks{Corresponding author: zhuang.z.ge@jyu.fi}
\affiliation{University of Jyvaskyla, Department of Physics, Accelerator laboratory, P.O. Box 35(YFL) FI-40014 University of Jyvaskyla, Finland}%
\author{Tommi~Eronen}
\affiliation{University of Jyvaskyla, Department of Physics, Accelerator laboratory, P.O. Box 35(YFL) FI-40014 University of Jyvaskyla, Finland}%
\author{Vasile~Alin~Sevestrean}\thanks{Corresponding author: sevestrean.alin@theory.nipne.ro}
\affiliation{International Centre for Advanced Training and Research in Physics, P.O. Box MG12, 077125 Bucharest-M\u{a}gurele, Romania}%
\affiliation{Faculty of Physics, University of Bucharest, 405 Atomiștilor, P.O. Box MG11, 077125 Bucharest-M\u{a}gurele, Romania}%
\affiliation{“Horia Hulubei” National Institute of Physics and Nuclear Engineering, 30 Reactorului, POB MG-6, RO-077125 Bucharest-M\u{a}gurele, Romania}
\author{Ovidiu~Ni\c{t}escu}
\affiliation{“Horia Hulubei” National Institute of Physics and Nuclear Engineering, 30 Reactorului, POB MG-6, RO-077125 Bucharest-M\u{a}gurele, Romania}
\author{Sabin~Stoica}
\affiliation{International Centre for Advanced Training and Research in Physics, P.O. Box MG12, 077125 Bucharest-M\u{a}gurele, Romania}%
\author{Marlom~Ramalho}
\affiliation{University of Jyvaskyla, Department of Physics, Accelerator laboratory, P.O. Box 35(YFL) FI-40014 University of Jyvaskyla, Finland}%
\author{Jouni~Suhonen}\thanks{Corresponding author:  jouni.t.suhonen@jyu.fi}%
\affiliation{University of Jyvaskyla, Department of Physics, Accelerator laboratory, P.O. Box 35(YFL) FI-40014 University of Jyvaskyla, Finland}%
\affiliation{International Centre for Advanced Training and Research in Physics, P.O. Box MG12, 077125 Bucharest-M\u{a}gurele, Romania}%
\author{Anu~Kankainen}
\affiliation{University of Jyvaskyla, Department of Physics, Accelerator laboratory, P.O. Box 35(YFL) FI-40014 University of Jyvaskyla, Finland}%
\author{Marjut~Hukkanen}
\affiliation{University of Jyvaskyla, Department of Physics, Accelerator laboratory, P.O. Box 35(YFL) FI-40014 University of Jyvaskyla, Finland}%
\affiliation{Centre d'Etudes Nucl\'eaires de Bordeaux Gradignan, UMR 5797 CNRS/IN2P3 - Universit\'e de Bordeaux, 19 Chemin du Solarium, CS 10120, F-33175 Gradignan Cedex, France}
\author{Arthur~Jaries}
\altaffiliation{Present address: Max-Planck-Institut f\"{u}r Kernphysik, Saupfercheckweg 1, 69117 Heidelberg, Germany}
\altaffiliation{Experimental Physics Department, CERN, CH-1211 Geneva 23, Switzerland}
\affiliation{University of Jyvaskyla, Department of Physics, Accelerator laboratory, P.O. Box 35(YFL) FI-40014 University of Jyvaskyla, Finland}%
\author{Ari~Jokinen} 
\affiliation{University of Jyvaskyla, Department of Physics, Accelerator laboratory, P.O. Box 35(YFL) FI-40014 University of Jyvaskyla, Finland}%
\author{Joel~Kostensalo}
\affiliation{Natural Resources Institute Finland, Yliopistokatu 6B, FI-80100, Joensuu, Finland}%
\author{Jenni~Kotila}
\affiliation{Finnish Institute for Educational Research, University of Jyv\"askyl\"a, P.O. Box 35, FI-40014, Jyv\"askyl\"a, Finland}%
\affiliation{Center for Theoretical Physics, Sloane Physics Laboratory Yale University, New Haven, Connecticut 06520-8120, USA}
\author{Maxime~Mougeot}
\affiliation{University of Jyvaskyla, Department of Physics, Accelerator laboratory, P.O. Box 35(YFL) FI-40014 University of Jyvaskyla, Finland}
\author{Iain~D.~Moore}
\affiliation{University of Jyvaskyla, Department of Physics, Accelerator laboratory, P.O. Box 35(YFL) FI-40014 University of Jyvaskyla, Finland}%
\author{Wirunchana~Rattanasakuldilok}
\affiliation{University of Jyvaskyla, Department of Physics, Accelerator laboratory, P.O. Box 35(YFL) FI-40014 University of Jyvaskyla, Finland}
\author{Jouni~Ruotsalainen}
\altaffiliation{Present address: GSI Helmholtzzentrum f\"ur Schwerionenforschung GmbH, 64291 Darmstadt, Germany}
\affiliation{University of Jyvaskyla, Department of Physics, Accelerator laboratory, P.O. Box 35(YFL) FI-40014 University of Jyvaskyla, Finland}
\author{Marek~Stryjczyk}
\altaffiliation{Present address: Institut Laue-Langevin (ILL), 71 Av. des Martyrs, 38000 Grenoble, France}
\affiliation{University of Jyvaskyla, Department of Physics, Accelerator laboratory, P.O. Box 35(YFL) FI-40014 University of Jyvaskyla, Finland}%
\date{\today}
\begin{abstract}
A high-precision measurement of the electron-capture (EC) decay $Q$ value for the ground-state-to-ground-state (gs-to-gs) transition of $^{113}$Sn to $^{113}$In has been performed using the JYFLTRAP double Penning trap mass spectrometer.
Employing the phase-imaging ion-cyclotron-resonance technique, the isomeric state of $^{113}$Sn at 77.389(19) keV was resolved, 
and the cyclotron frequency ratio measured between the isomer $^{113m}$Sn and the daughter nucleus $^{113}$In. This yielded an isomer-to-ground-state $Q$ value of 1116.64(19) keV and gs-to-gs $Q$ value of  1039.25(19) keV. 
The atomic mass excess of $^{113}$Sn was determined as $-$88327.87(27) keV/c$^2$,  in excellent agreement with the Atomic Mass Evaluation 2020 (AME2020) but with a sixfold precision improvement. 
Using nuclear energy-level data for $^{113}$In, we identified two low $Q$-value transitions of the ground state of $^{113}$Sn to  excited states of $^{113}$In at 1024.280(50) keV ($Q_{EC}^* = 14.97(20)$ keV, second forbidden non-unique) and 1029.650(50) keV ($Q_{EC}^* = 9.60(20)$ keV, allowed). The allowed transition exhibits small energy differences ($\Delta_{L1} = 5.58(20)$ keV, $\Delta_{L2} = 5.87(20)$ keV) from L1 and L2 shell binding energies, enhancing endpoint events. 
Partial half-lives and energy-release spectra were calculated using the self-consistent Dirac–Hartree–Fock–Slater (DHFS) method (including exchange, overlap, shake-up, and shake-off corrections) together with the nuclear shell model,
show enhanced endpoint sensitivity for the allowed transition to the state at 1029.650 keV. Including subthreshold atomic states in the spectral function enhances the EC rate near the zero-neutrino-momentum region by a factor of five, enabling new approaches for low $Q$-value EC reactions in neutrino-mass studies.
\end{abstract}
\maketitle
\section{Introduction}

Neutrino oscillations, firmly established by solar, atmospheric, and accelerator experiments, prove that neutrinos have mass and the weak flavor eigenstates are superpositions of three mass eigenstates~\cite{Fukuda1998,SNOCollaboration2002,Planck2018,Adame2025}.
These oscillations provide only squared mass differences, leaving the absolute neutrino mass scale undetermined~\cite{Gerbino2018a}. 
The absolute mass hierarchy and scale remain elusive, critical for understanding leptogenesis, dark matter, and cosmic evolution~\cite{Planck2018, Adame2025, Giusarma2023}. Particle physics demands direct, model-independent probes to resolve the Dirac/Majorana nature without relying on potentially incorrect theoretical assumptions inherent in specific beyond-Standard-Model extensions.

Direct kinematic measurements of weak nuclear decays offer a model-independent approach to probe the absolute neutrino mass by analyzing energy-momentum conservation without direct neutrino detection~\cite{Drexlin2013}. These inquiries constrain the effective electron neutrino or antineutrino mass, an essential metric in nuclear physics, particle physics, and cosmology~\cite{Avignone2008,Ejiri2019}. The KArlsruhe TRItium Neutrino (KATRIN) experiment provides the leading direct  antineutrino-mass evaluations via examination of the tritium $\beta$-decay spectrum, recently establishing a threshold of $m_{\overline{\nu}_e} < 0.45$ eV/$c^2$ (90\% C.L.) and expects to reach a
 sensitivity of better than 0.3 eV/$c^2$ (90\% C.L.) by the end of year 2025~\cite{KATRIN2025}. Project 8 leverages cyclotron radiation emission spectroscopy on tritium, yielding a preliminary bound of $m_{\overline{\nu}_e}<155$ eV/$c^2$ with a prospective sensitivity of 0.04 eV/$c^2$~\cite{Ashtar23}. Conversely, calorimetric methodologies, such as the HOLMES project employing $^{163}$Ho electron capture (EC), have instituted a threshold of $m_{\nu_e} < 27$ eV/$c^2$~\cite{HOLMES2025}, whereas the ECHo initiative has attained <15 eV/$c^2$~\cite{ECHo2025}.

Low $Q$-value transitions enhance the fraction of decays near the spectral endpoint, improving sensitivity to neutrino mass effects in single-decay experiments~\cite{Ferri2015}. Currently, ground-state-to-ground-state (gs-to-gs) low-$Q$-value decays of $^3$H and $^{163}$Ho are utilized for direct (anti)neutrino-mass determination. 
A new frontier lies in ground-state-to-excited-state (gs-to-es) transitions with $Q \lesssim 10$ keV, where $\beta^-$ endpoint rates rise steeply and EC atomic resonance tails enhance spectral features~\cite{Mustonen2010,Mustonen2011,Haaranen2013,Suhonen2014,Sandler2019,Karthein2019a,DeRoubin2020,ge2021,ge2021b,Ge2022a,ERONEN2022,Ge2022b,Ramalho2022,Gamage22,Keblbeck2023,Ge2023,valverde2024,Ge2024,Ge2024b,ge2025a}.
Ongoing efforts with Penning trap mass spectrometry (PTMS) at radioactive beam facilities like JYFLTRAP, LEBIT, CPT, ISOLTRAP, and SHIPTRAP focus on identifying isotopes with low $Q$-value gs-to-es $\beta^{\pm}$/EC transitions suitable for such experiments~\cite{Sandler2019,Karthein2019a,DeRoubin2020,ge2021,ge2021b,Ge2022a,ERONEN2022,Ge2022b,Ramalho2022,Gamage22,Keblbeck2023,Ge2023,valverde2024,Ge2024,Ge2024b,ge2025a,Nesterenko2014,Eliseev2015}. PTMS delivers sub-keV $Q$-value precision and remains the only direct method capable of verifying ultra-low (<1 keV) $Q$-value transitions. 
Recent breakthroughs in low gs-to-es ($Q^*$) transitions (especially ultra-low) include the allowed Gamow-Teller EC in $^{159}$Dy(3/2$^-$) $\to$ $^{159}$Tb$^*$(5/2$^-$) at 363.5449~keV with $Q^*_{EC}=1.18(19)$~keV~\cite{ge2021b}, recently further recognized as a sub-eV neutrino-mass candidate~\cite{FU2025,Kishimoto2025}; in $\beta^-$-decay, the second-forbidden unique $^{115}$In(9/2$^+_{\rm gs}$) $\to$ $^{115}$Sn$^*$(3/2$^+$) is experimentally confirmed as the lowest $Q$-value case with $Q^*_\beta = 147(10) 
$~eV~\cite{Wieslander2009,Mount2009,Andreotti2011,FORMAGGIO2021,Zheltonozhsky2018}, while $^{135}$Cs(7/2$^+$) $\to$ $^{135}$Ba$^*$(11/2$^-$) gives $Q^*_\beta=440(310)$~eV (first-forbidden unique)~\cite{DeRoubin2020}, $^{131}$I(7/2$^+$) $\to$ $^{131}$Xe$^*$(9/2$^+$) yields $Q^*_\beta=1.03(23)$~keV (allowed)~\cite{ERONEN2022}, and the $\beta$-decaying isomer $^{110m}$Ag(6$^+$) $\to$ $^{110}$Cd$^*$(5$^+$) holds the lowest allowed $\beta^-$-decay $Q^*$ at $405(135)$~eV~\cite{Karthein2019a}. EC cases are particularly promising when $Q^*$ aligns with atomic shell binding energies to produce resonant enhancement near the endpoint to show an ultra-low energy difference between $Q^*$ and the atomic relaxation energy~\cite{ge2021b,Ge2022a,Ge2024b}; among them, $^{159}$Dy is exceptional, while $^{111}$In(9/2$^+$) $\to$ $^{111}$Cd$^*$(7/2$^+$, 853.94(7) keV)~\cite{Ge2022a} and $^{95}$Tc(9/2$^+$) $\to$ $^{95}$Mo$^*$ (9/2$^+$, 1675.40(60) keV)~\cite{Ge2024b} show strong resonance potential as good candidates for future neutrino-mass experiments.

Here, we report the high-precision gs-to-gs EC $Q$-value measurement for $^{113}$Sn using the JYFLTRAP PTMS. By combining this precise $Q$ value with nuclear energy-level data for the excited states of the daughter nucleus $^{113}$In, we evaluate potential gs-to-es $Q$ values. Two low $Q$-value transitions in $^{113}$Sn are identified as promising candidates for neutrino-mass determination, including an allowed transition at $Q$ = 9.60(20) keV with L-shell proximity and subthreshold atomic enhancement~\cite{Kishimoto2025}. Theoretical predictions for partial half-lives and energy-release distributions of these transitions were obtained using the self-consistent many-electron Dirac--Hartree--Fock--Slater (DHFS) method and the nuclear shell model, accounting for key atomic and nuclear corrections.

\section{Experimental method}
\begin{figure}[!htb]
\centering
\includegraphics[width=0.95\columnwidth]{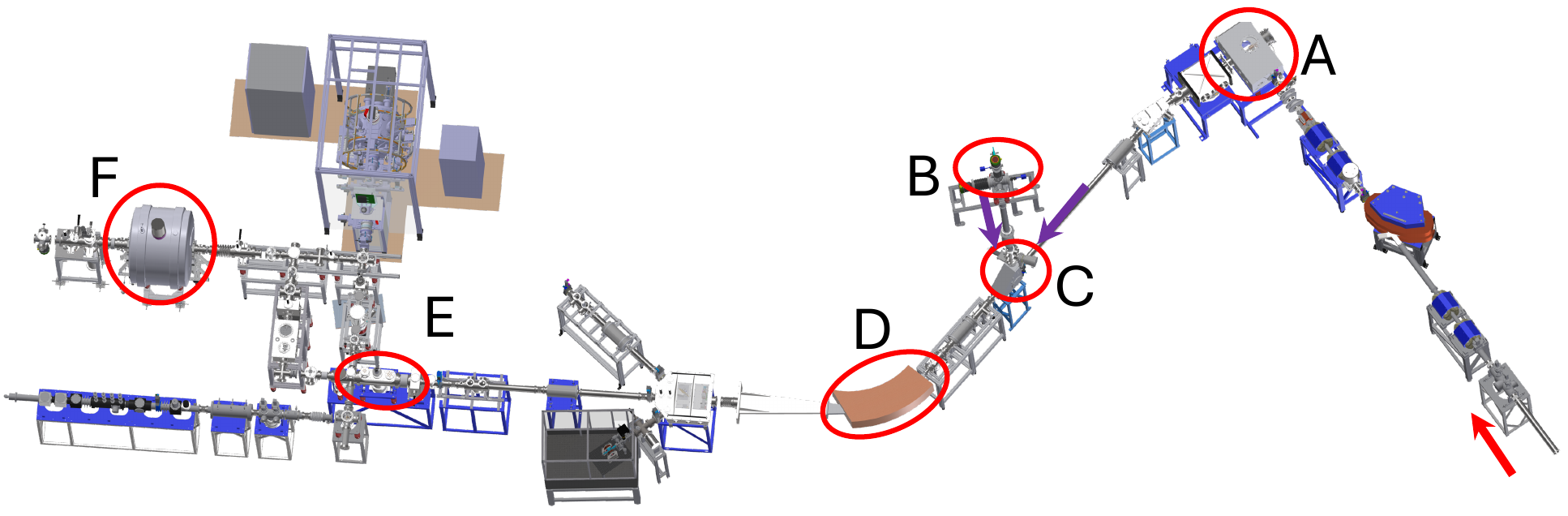}
\caption{(Color online).  Schematic view of the IGISOL facility and the experimental setup. Mass $A$ = 113 ions (including $^{113}$Sn$^{+}$, $^{113m}$Sn$^{+}$, $^{113}$Cd$^{+}$ and $^{113}$In$^{+}$) were generated via $\alpha$-induced fusion-evaporation reactions on a natural Cd target in the IGISOL target chamber (A). 
These ions were selected by the electrostatic kicker (C) and the dipole magnet (D), and then cooled and bunched in the RFQ-CB (E). The final $Q$ value and mass measurements were performed using the JYFLTRAP Penning-trap setup (F). While the offline ion source (B) can produce stable $^{113}$In$^{+}$ ions using glow discharge, the reference ions for this specific experiment were sourced directly from the target chamber.
}
\label{fig:igisol}
\end{figure}

The experiment was conducted at the Ion Guide Isotope Separator On-Line (IGISOL) facility using the JYFLTRAP double Penning trap mass spectrometer at the University of Jyv\"askyl\"a, Finland~\cite{Eronen2012,Moore2013,Kolhinen2013}. The measurement targeted the direct EC decay $Q$ value measurements of $^{113}$Sn, utilizing a natural cadmium (Cd) target bombarded with a 50 MeV $\alpha$ beam from the K-130 cyclotron to produce ions of interest via fusion-evaporation reactions.
The produced radioactive ions from the reaction were stopped in a gas-filled light ion guide, and the ions were extracted using helium gas flow and electric fields through a sextupole ion guide~\cite{Karvonen2008} as shown in Fig.~\ref{fig:igisol}. The ions were then accelerated to 30 kV and subjected to mass separation using a 55$^\circ$ dipole magnet with a mass resolving power of $M/\Delta M \approx 500$. For ions with A/q = 113, including $^{113}$In$^{+}$, $^{113m}$Sn$^{+}$, and $^{113}$Sn$^{+}$, cooling and bunching were performed in a radiofrequency-quadrupole cooler-buncher (RFQ-CB), as described in ~\cite{Nieminen2001,VIRTANEN2025}. 

The JYFLTRAP system consists of two cylindrical Penning traps within a 7-T superconducting solenoid. The first trap, filled with helium gas, serves as a purification trap, achieving isobaric purification with a resolving power of $\approx 10^5$ via the sideband buffer gas cooling technique~\cite{Savard1991}. All ions, including $^{113}$In$^{+}$, $^{113m}$Sn$^{+}$, $^{113}$Sn$^{+}$, and $^{113}$Cd$^{+}$, were excited to a large magnetron orbit using a dipole excitation at the magnetron frequency $\nu_-$ for 11 ms, followed by quadrupole excitation for 100 ms to center the desired ions through buffer gas collisions. 
This method suppresses most contaminants. 
%
Subsequently, the beam was transferred to the second (precision) trap, where the cyclotron frequencies of cleaned ion samples of $^{113m}$Sn$^{+}$ and $^{113}$In$^{+}$ (from which the isomer‑to‑ground‑state $Q$ value is derived) were measured using the phase‑imaging ion‑cyclotron‑resonance (PI‑ICR) technique~\cite{Nesterenko2018, nesterenko2021}.

\begin{figure}[!htb]
\includegraphics[width=0.9\columnwidth]{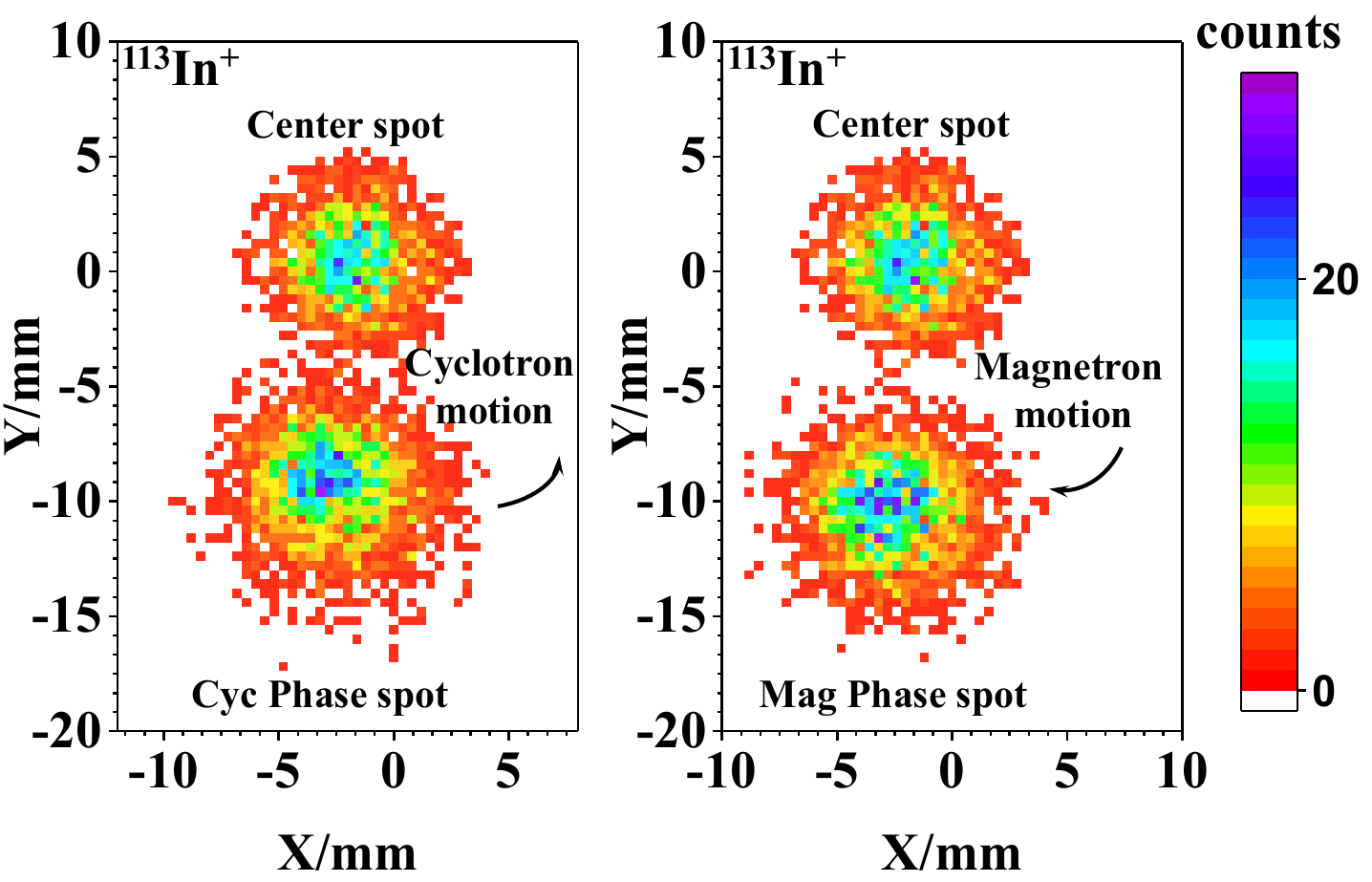}
\caption{(Color online). Example $^{113}$In$^{+}$ ion spots of center, cyclotron phase, and magnetron phase on the 2-dimensional position-sensitive MCP detector after a typical PI-ICR excitation pattern with an accumulation time of 738 ms. The magnetron phase spot is shown on the left and the cyclotron phase spot on the right. The angle difference between these spots relative to the center is used to infer the cyclotron frequency. The color in each pixel corresponds to the number of ions, represented by the color bars.}
   \label{fig:2-phases}
\end{figure}

For $Q$-value measurements, the PI-ICR method was used to measure the cyclotron frequency, $\nu_c = qB/(2\pi m)$, where $B$ is the magnetic field strength, $q$ is the ion charge, and $m$ is the ion mass. 
Two timing patterns, differing in the timing of a quadrupolar conversion pulse with a phase-accumulation time $t_{acc}$, were used to project ions onto a position-sensitive MCP detector, accumulating free evolutions of the magnetron and reduced cyclotron phases in the precision trap. A center spot was recorded without excitations to determine the phase angles, as shown in Fig.~\ref{fig:2-phases} for $^{113}$In$^{+}$.
The cyclotron frequency is derived from the angle between the magnetron and cyclotron phase images relative to the center spot, $\alpha_c = \alpha_+ - \alpha_-$, using:
\begin{equation}
\label{eq:nuc2}
\nu_c = \frac{\alpha_c + 2\pi n_c}{2\pi t_{acc}},
\end{equation}
where $n_c$ is the number of full revolutions during $t_{acc}$. Multiple accumulation times for $^{113m}$Sn$^{+}$ ensured an unambiguous $n_c$ assignment, with 244 ms and 738 ms used for final $\nu_c$ measurements of $^{113m}$Sn$^{+}$ and $^{113}$In$^{+}$, preventing overlap with impurities.
Phase spots were positioned to keep the cyclotron phase angle $\alpha_c$ within a few degrees, minimizing systematic errors in the cyclotron frequency ratio to below $10^{-10}$~\cite{Eliseev2014}.
Figure~\ref{fig:2-phases} depicts a representative measurement with cyclotron and magnetron phase spots relative to the central spot.
Delays were scanned over magnetron and cyclotron periods to correct for residual motion impacts, ensuring accurate phase measurements. The total interleaved data accumulation time for $\nu_c$ determinations of $^{113m}$Sn$^{+}$-$^{113}$In$^{+}$ approximated 8.5 hours.

The isomeric EC $Q$ value, $Q_{EC}^{m}$, was calculated from the mass difference:
\begin{equation}
\label{eq:Qec}
Q_{EC}^{m} = (M_p - M_d)c^2 = (R-1)(M_d - qm_e)c^2 + (R \cdot B_d - B_p),
\end{equation}
where $M_p$ and $M_d$ are the masses of the parent ($^{113m}$Sn) and daughter ($^{113}$In) atoms, respectively, $R = \nu_{c,d}/\nu_{c,p}$ is the cyclotron frequency ratio for singly charged ions ($q=1$), and $m_e$ is the electron mass. 
Electron binding energies $B_p$ and $B_d$ are disregarded owing to their minimal magnitudes (a few eV~\cite{NIST_ASD}), with $R \approx 1$.

The gs-to-gs $Q$ value, $Q_{EC}$, was derived by combining $Q_{EC}^{m}$ with the isomeric excitation energy of $^{113m}$Sn, 77.389(19) keV, as documented by the National Nuclear Data Center~\cite{NNDC}:
\begin{equation}
\label{eq:Qec_gs}
Q_{EC} = Q_{EC}^{m} - E^{*}_m,
\end{equation}
where $E^{*}_m$ is the isomeric excitation energy. The identical $A/q$ and small relative mass difference ($\Delta M/M < 10^{-4}$) render mass-dependent errors negligible relative to statistical uncertainties. 
The mass uncertainty of the reference ($^{113}$In, 0.19 keV/$c^2$~\cite{Wang2021}) likewise exerts negligible influence on the $Q$ value.

\begin{table*}[!htb]
\caption{Final results for the mean cyclotron frequency ratio between the ground state of daughter $^{113}$In and parent ($^{113m}$Sn, 7/2$^{+}$) and ground state ($^{113}$Sn, 1/2$^{+}$) nuclei. The frequency ratio $\overline{R}$ corresponds to the $^{113}$In$^{+}$/$^{113m}$Sn$^{+}$ pair, measured experimentally as 1.00001061759(18). The $Q_{EC}$ values (in keV) for the isomer-to-ground-state (is-to-gs) and gs-to-gs transitions, and the mass excess (in keV/$c^2$) for $^{113m}$Sn and $^{113}$Sn, are sourced from literature (Lit.)~\cite{Wang2021} and this work, 
respectively. The isomeric excitation energy of $^{113m}$Sn is 77.389(19) keV~\cite{NNDC}.}
\begin{ruledtabular}
\begin{tabular}{c c c c c c} 
\multirow{2}{*}{\centering Nucleus} & \multirow{2}{*}{\centering $\overline{R}$} & \multicolumn{2}{c}{$Q_{EC}$ (keV)} & \multicolumn{2}{c}{mass excess (keV/$c^2$)} \\
\cline{3-4}\cline{5-6} 
& & Lit. & This work & Lit. & This work \\
\hline\noalign{\smallskip}
$^{113m}$Sn & 1.0000106176(18) & 1116.4(16) & 1116.64(19) & -88250.7(16) & -88250.48(27) \\
$^{113}$Sn & & 1039.0(16) & 1039.25(19) & -88328.1(16) & -88327.87(27) \\
\end{tabular}
\end{ruledtabular}
\label{table:Q-value}
\end{table*}

\begin{table*}[!htb]
\caption{Low $Q$-value candidate transitions from the ground state of the parent nucleus $^{113}$Sn (1/2$^{+}$) to the excited states of the daughter nucleus $^{113}$In. The first column lists the excited final state of $^{113}$In with the experimentally known spin-parity assignment indicated. The decay type is provided in the second column. The third and fourth columns present the gs-to-gs $Q_{EC}$ values, sourced from literature (Lit.)~\cite{Wang2021} and this work, respectively. The fifth and sixth columns show the gs-to-es $Q_{EC}^*$ values. The seventh column displays the experimental excitation energy $E^{*}$ with its error~\cite{NNDC}. The eighth column shows the confidence ($\sigma$) of $Q_{EC}^*$ being positive/negative. Columns nine to eleven, denoted as $\Delta_x$, represent the distance of $Q_{EC}^*$ values to the binding energy $\varepsilon_x$ of electrons in the daughter atoms, sourced from~\cite{X-Ray_Data_Booklet}. All units for energy and $Q$ values are in keV.
FNU denotes forbidden non-unique.
}
\begin{ruledtabular}
\begin{tabular*}{\textwidth}{@{\extracolsep{\fill}}cc cc cc c c cc c@{\extracolsep{\fill}}}
\multirow{2}{*}{\centering Final state} & \multirow{2}{*}{\centering Decay type} & \multicolumn{2}{c}{$Q_{EC}$} & \multicolumn{2}{c}{$Q_{EC}^*$} & \multirow{2}{*}{\centering E$^{*}$} & \multirow{2}{*}{\centering $Q/\delta Q$} & \multicolumn{3}{c}{$\Delta_x$ (This work)} \\
\cline{3-4}\cline{5-6}\cline{9-11}
& & Lit. & This work & Lit. & This work & & & L1 & L2 & M1 \\
\hline\noalign{\smallskip}
$^{113}$In(5/2$^{+}$)& 2nd FNU & 1039.0(16) & 1039.25(19) & 14.7(16) & 14.97(20) & 1024.280(50) & 76 & 10.95(20) & 11.24(20) & 14.20(20) \\
$^{113}$In(1/2$^{+}$, 3/2$^{+}$) & Allowed & 1039.0(16) & 1039.25(19) & 9.4(16) & 9.60(20) & 1029.650(50) & 49 & 5.58(20) & 5.87(20) & 8.83(20) \\
\end{tabular*}
\end{ruledtabular}
\label{table:low-Q}
\end{table*}

\section{Results and discussion}
The isomeric $Q$ value, $Q_{EC}^{m}$, was determined using the cyclotron frequency ratio $R$ via Eq.~\eqref{eq:Qec}. Two data sets for $^{113m}$Sn$^{+}$-$^{113}$In$^{+}$ were collected using PI-ICR. Each measurement cycle, including magnetron phase, cyclotron phase, and center spot scans, was completed in under 5 minutes for the decay pair.
Spot positions were fitted using the maximum likelihood method, summing cycles for adequate statistics. Phase angles were computed to infer cyclotron frequencies. The reference $\nu_c$ for $^{113}$In$^{+}$ was linearly interpolated to the $^{113m}$Sn$^{+}$ measurement time to compute $R$. 
Solely bunches with fewer than three ions were scrutinized to mitigate frequency shifts from ion-ion interactions, as discussed in~\cite{Kellerbauer2003,Roux2013,nesterenko2021}, with no count-rate-related shifts observed.
Temporal magnetic field variations, $\delta_B(\nu_c)/\nu_c = \Delta t \times 2.01(25) \times 10^{-12}$/min~\cite{nesterenko2021}, were considered, with $\Delta t < 10$ minutes assuring uncertainties below $10^{-10}$. 
Frequency shifts from ion image distortions were negligible, and mass doublet properties of $^{113m}$Sn$^{+}$-$^{113}$In$^{+}$ canceled systematic uncertainties.

\begin{figure}[!htb]
\includegraphics[width=0.99\columnwidth]{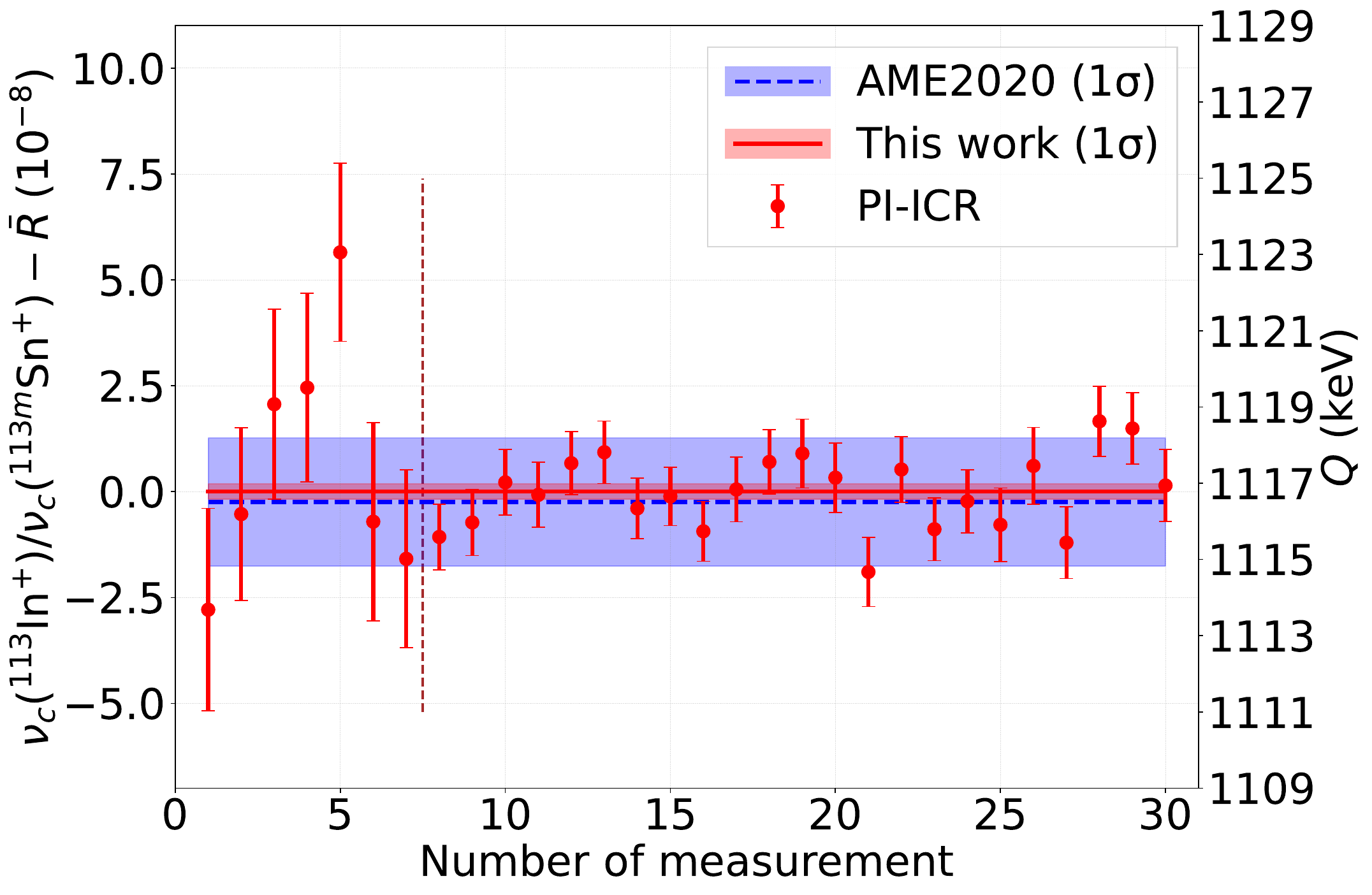}
\caption{(Color online). Measured experimental results compared to literature values~\cite{Huang2021,Wang2021}. Deviations of individual cyclotron frequency ratios $R$ ($\nu_c(^{113}$In$^{+})/\nu_c(^{113m}$Sn$^{+}$)) from the mean $\overline{R}$ (left axis) and $Q$ value (right axis) are shown relative to AME2020 values. Red points with uncertainties represent PI-ICR measurements. 
The weighted average $\overline{R}$ (Table~\ref{table:Q-value}) is shown as a solid red line with a 1$\sigma$ uncertainty band in red. The dashed blue line indicates the difference from AME2020, with its 1$\sigma$ uncertainty area shaded in blue. 
Two data sets with accumulation times of 244~ms (left) and 738~ms (right) are presented,  with the individual cyclotron frequency ratios separated by a dashed brown line.
}
   \label{fig:ratio}
\end{figure}

The weighted mean ratio $\overline{R}$ was calculated with inner and outer errors to ascertain the Birge ratio~\cite{Birge1932}, utilizing the maximum as the weight. Figure~\ref{fig:ratio} presents the outcomes compared to literature values. The ultimate $\overline{R}$ is 1.00001061759(18), yielding $Q_{EC}^{m} = 1116.64(19)$ keV and $Q_{EC} = 1039.25(19)$ keV using Eq.~\ref{eq:Qec_gs} with $E^{*}_m = 77.389(19)$ keV. Table~\ref{table:Q-value} contrasts results with AME2020.

The gs-to-gs $Q_{EC}$ value of 1039.25(19) keV is approximately eight times more precise than the AME2020 value, with a deviation of 0.3(16) keV, consistent within 1$\sigma$. The AME2020 $Q_{EC}$ is primarily derived from $\beta^-$-decay experiments~\cite{Wang2021}. Employing the $^{113}$In mass excess of $-$90212.74(19) keV/c$^2$ from AME2020~\cite{Wang2021}, the $^{113}$Sn (1/2$^{+}$) mass excess is $-$88327.87(27) keV/$c^2$, with a half-life of 115.09(3) days, as reported by the National Nuclear Data Center~\cite{NNDC}, enhancing precision by a factor of six. The $^{113m}$Sn mass excess is deduced to be $-$88250.48(27) keV/$c^2$, also showing improved precision.

The precise $Q_{EC}$ value from this work, combined with nuclear energy-level data for $^{113}$In excited states from the National Nuclear Data Center~\cite{NNDC}, resulted in the determination of gs-to-es $Q$ values ($Q_{EC}^*$) for two states, confirming that they are energetically allowed (Table~\ref{table:low-Q}). In EC decays, proximity of the $Q$ value to electron ionization energies enhances endpoint event rates, critical for neutrino mass sensitivity~\cite{Ferri2015}. The $Q_{EC}^*$ dependence is steeper than in $\beta^-$-decay.

Table~\ref{table:low-Q} lists $\Delta_x$, the difference between $Q_{EC}^*$ and electron binding energies $\varepsilon_x$ for the most favorable allowed atomic shells L1, L2, and M1
of the daughter nucleus $^{113}$In~\cite{X-Ray_Data_Booklet}. For the 1024.280(50) keV state (5/2$^{+}$), electron captures from L1, L2, L3, M1, M2, and higher shells are allowed for the second forbidden non-unique transition from $^{113}$Sn (1/2$^{+}$). For the 1029.650(50) keV state, only s- and p$_{1/2}$-electrons (e.g., L1, L2, M1, M2) are permitted due to angular momentum conservation in this allowed transition. The $^{113}$Sn (1/2$^{+}$) $\rightarrow$ $^{113}$In$^*$ (1029.650 keV) transition yields $\Delta_{\rm L1} = 5.58(20)$ keV and $\Delta_{\rm L2} = 5.87(20)$ keV, relative to the binding energies of the electrons,  $\varepsilon_{\rm L1} = 4.02$ keV and $\varepsilon_{\rm L2} = 3.73$ keV~\cite{X-Ray_Data_Booklet}. As an allowed transition with a half-life of $20^{+2.3}_{-2.0} \times 10^9$ yr (Table~\ref{table:T12-allowed}), it enhances endpoint events (see Figs.~\ref{fig:4-spectra}, \ref{fig:5-zoom-spectra}).
This enhancement—driven by phase-space scaling $\propto  (1/\Delta_x)^2$ and atomic effects—positions $^{113}$Sn as a complementary candidate for future neutrino-mass experiments, despite its overall EC rate being lower than for $^{163}$Ho. Its moderate half-life (115.09(3) d) and clean $\gamma$-tagging via the 1029.650 keV cascade enable scalable source production and background rejection in cryogenic microcalorimeters, offering a possible pathway to sub-eV sensitivity alongside  $^{163}$Ho and $^{159}$Dy. However, further precise measurement of the branching ratio of this transition is required.


\section{Theoretical predictions}
To predict transition half-lives and energy-release distributions, we employed the atomic self-consistent many-electron Dirac--Hartree--Fock--Slater (DHFS) method and nuclear shell model, proven effective in prior work~\cite{SevestreanPRA2023}. We used the DHFS method to compute electron wave functions and energy levels for initial and final atoms, with the initial atom in its ground state and the final atom having a hole in each atomic shell allowing an electron to be captured. A central field potential $V(r)$ was used, as described by Salvat et al.~\cite{SalvatCPC2019}:
\begin{equation}
V_{\mathrm{DHFS}}(r) = V_{\mathrm{nuc}}(r) + V_{\mathrm{el}}(r) + V_{\mathrm{ex}}(r),
\end{equation}
with nuclear, electronic, and exchange components. The nuclear potential accounts for a Fermi charge distribution~\cite{HahnPR1956}, the electronic potential reflects the electron cloud, and the exchange potential ensures correct asymptotic behavior with latter tail correction. Our calculations employed the RADIAL package~\cite{SalvatCPC2019}.

The electron bound states for the initial system 
are denoted $\ket{(n, \kappa)}$, with $n$ the principal quantum number and $\kappa$ the relativistic quantum number. Ionization energies $\varepsilon_x$ were theoretically computed using the DHFS routine in the RADIAL package.  
For allowed transitions, the energy distribution of an EC event is:
\begin{equation}
\label{eq:rho}
\rho(E) = \frac{G_{\beta}^{2}}{(2 \pi)^{2}} C \sum_{x} n_{x} \beta_{x}^{2} B_{x} S_{x} p_{\nu} E_{\nu} \frac{\Gamma_{x} /(2 \pi)}{\left(E-\varepsilon_{x}\right)^{2}+\Gamma_{x}^{2}/4},
\end{equation}
where $B_x$ includes the exchange and overlap corrections, and $S_x$ includes the shake-up and shake-off correction.
$E = Q_{EC}^* - E_{\nu}$, $p_{\nu} = \sqrt{E_{\nu}^2 - m_{\beta}^2}$, $\beta_x$ is the Coulomb amplitude, $n_x$ is shell occupancy, and $\Gamma_x$ is the Breit-Wigner resonance width~\cite{CampbellADNDT2001,RanitzschARXIV2014}. The Fermi constant and Cabibbo angle combine as $G_\beta = G_F \cos \theta_C$. In the energy distribution the sum over x takes into account the contribution from subthreshold states as described in \cite{Kishimoto2025}. The final shape in the region of interest, where the neutrino-mass impact can be observed, takes into account contributions from both the subshells that are energetically available and from the subshells that have the central value of energy under the threshold but a width extending by a significant amount to the region of interest. Thus, additional subshells can contribute to the energy-density distribution. The shape factor $C$ for allowed transitions includes the nuclear form factor~\cite{Behrens1982}:
\begin{equation}
C = \left[{ }^{A} F_{101}^{(0)}\right]^{2} = \left[-\frac{g_{A}}{\sqrt{2 J_{i}+1}} M_{\mathrm{GT}}\right]^{2},
\end{equation}
with $M_{\mathrm{GT}}$ the Gamow--Teller matrix element~\cite{JSuhonen2007}, $J_i$ the initial nuclear angular momentum, and value of the axial coupling taken to be $g_A = 0.7$, as suggested by the $g_{\rm A}$ systematics advocated in \cite{Suhonen2017}.

The  decay probability in the energy interval $(0,E)$ is:
\begin{equation}
\lambda(E) = \int_{0}^{E} \rho(E') dE',
\end{equation}
with the total decay constant being $\lambda = \lambda(Q_{EC}^* - m_\nu)$, which can be written as:
\begin{equation}
\lambda = \sum_{x} \lambda_x,
\end{equation}
where,
\begin{equation}
\lambda_x = \frac{G_{\beta}^{2}}{(2 \pi)^{2}} C n_x \beta_x^{2} B_x S_x p_{\nu}(Q_{EC}^{i} - \varepsilon_x).
\end{equation}

In the calculation of the decay constant for the forbidden non-unique transition, the shape factor C has a more complex formula as described in \cite{BambynekRMP1977}.

\subsection{Exchange and overlap corrections}
In the lowest-order EC process, an electron is captured from shell $x$, leaving a hole. Due to electrons being indistinguishable, a higher-order process may involve capture from shell $y$ with an electron from $x$ promoted to $y$, leaving a hole in $x$. This exchange correction is accounted for as~\cite{MougeotARI2018}:
\begin{equation}
B_{n \kappa} = \left|\frac{b_{n \kappa}}{\beta_{n \kappa}}\right|^{2},
\end{equation}
where,
\begin{eqnarray}
b_{n \kappa} &=& \left[\prod_{m, \mu} \bra{(m, \mu)'}\ket{(m, \mu)}^{n_{m \mu}}\right] \bra{(n, \kappa)'}\ket{(n, \kappa)}^{-\frac{1}{2|\kappa|}} \nonumber \\
&& \times \left[\beta_{n \kappa} - \sum_{m \neq n} \beta_{m \kappa} \frac{\bra{(m, \kappa)'}\ket{(n, \kappa)}}{\bra{(m, \kappa)'}\ket{(m, \kappa)}}\right].
\end{eqnarray}
This overlap correction arises from imperfect overlap of initial $\ket{(m, \kappa)}$ and final $\ket{(m, \kappa)'}$ wave functions due to the sudden change in nuclear charge.

\begin{table*}[!htb]
\caption{Computed half-lives for the EC decay of $^{113}$Sn to the excited states in $^{113}$In, with the $Q_{EC}^*$ values shown in Table~\ref{table:low-Q}. 
The shell-model interaction used is jj45pnb.}
\begin{ruledtabular}
\begin{tabular*}{\textwidth}{lc|ccccccccccc}
$Q_{EC}^*$ & Final & Total half-life & $\mathrm{L1}$ & $\mathrm{L2}$ & $\mathrm{L3}$ & $\mathrm{M1}$ & $\mathrm{M2}$ & $\mathrm{M3}$ & $\mathrm{M4}$ & $\mathrm{N1}$ & $\mathrm{N2}$ \\
(keV) & J & (yr) & (yr) & (yr) & (yr) & (yr) & (yr) & (yr) & (yr) & (yr) & (yr) \\
\hline\noalign{\smallskip}
	9.60 & 3/2 & 33.56 & 59.60 & 2327.62 & - & 105.54 & 4158.28 & - & -  & 403.91 & 17428.1 \\
	9.60 & 1/2 & 238.08 & 422.75 & 16510.2 & - & 748.59 & 29495.3 & - & - & 2864.97 & 123620. \\
	14.97 & 5/2 & 4.04 $\times 10^{9}$ & 7.15 $\times 10^{11}$ & 
	2.79 $\times 10^{13}$ & 5.85 $\times 10^{9}$ & 
	1.12 $\times 10^{12}$ & 4.40 $\times 10^{13}$ & 
	1.65 $\times 10^{10}$ & 6.94 $\times 10^{12}$ & 
	4.13 $\times 10^{12}$ & 1.78 $\times 10^{14}$ \\

\end{tabular*}
\end{ruledtabular}
\label{table:T12-allowed}
\end{table*}

\subsection{Shake-up and shake-off effects}

During the EC decay, a spectator electron can be promoted to upper unoccupied shells, and this process is called the shake-up effect. The spectator electron can also be ejected to the continuum, and this process is called the shake-off effect. We consider a maximum of one spectator electron undergoing each of these transitions, such that in the final state, we would have at most two holes in the atomic shell. We use the formalism developed in \cite{MougeotARI2018} to end up with the following expression for the $S_{x}$ factor from Eq. \ref{eq:rho},

\begin{equation}
	S_{n \kappa}=1+\sum_{m, \mu}P_{m \mu} .
\end{equation}
Here $P_{m \mu}$ represents the probability that a spectator electron from  $\ket{(m \mu)}$ undergoes the shake-up or shake-off process. A spectator electron can remain in the initial shell, can swap shells with another spectator electron or be promoted or ejected. Thus $P_{m \mu}$ can be computed as unity minus the probability for the electron to remain in the initial shell and the probability to swap shells, leading to
\begin{eqnarray}
	P_{m \mu}=&&1-|\bra{(m, \mu)^{\prime}}\ket{(m, \mu)}|^{2 n_{m\mu}}\nonumber\\
	&&-\sum_{l \neq m} n_{l \mu}^{\prime} n_{m \mu}|\bra{(l, \mu)^{\prime}}\ket{(m, \mu)}|^2.
\end{eqnarray}

\begin{figure}[!htb]
\includegraphics[width=0.9\columnwidth]{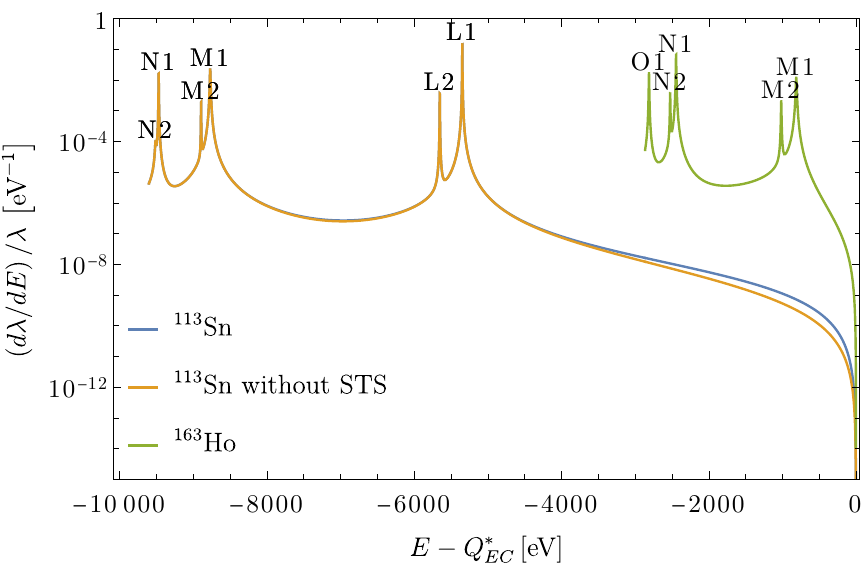}
   \caption{Normalized released-energy distribution for the allowed electron capture decay of $^{113}$Sn to the excited state of $^{113}$In ($J^\pi = 3/2^{+}$) shown in blue alongside the same transition without the subthreshold (STS) state with orange, in comparison with the electron capture decay of $^{163}$Ho shown in green. The peaks labeled K, L$_1$, L$_2$, M$_1$, M$_2$, N$_1$, N$_2$, and O$_1$ correspond to capture from the respective atomic subshells.}
   \label{fig:4-spectra}
\end{figure}

\begin{figure}[!htb]
   \includegraphics[width=0.9\columnwidth]{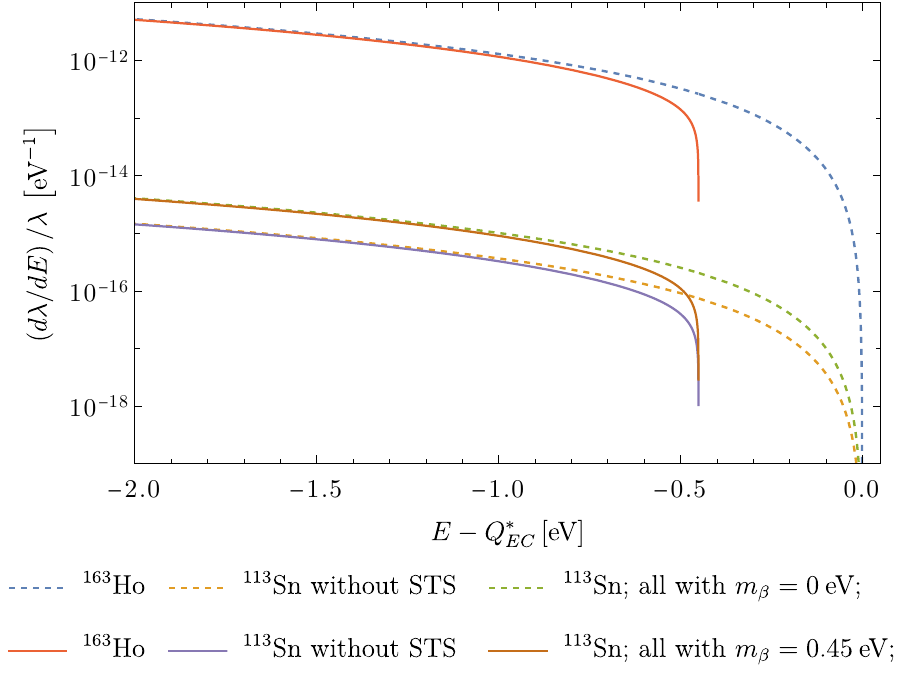}
   \caption{Enlarged detail of Fig.~\ref{fig:4-spectra} showing the effect of neutrino masses of 0.45 eV and 0 eV. Dotted lines represent spectra for a massless neutrino, while solid lines correspond to a 0.45 eV neutrino mass. 
   The corresponding lines for the holmium EC are also indicated for comparison. The curves without the subthreshold (STS) contribution are included for reference.}
   \label{fig:5-zoom-spectra}
\end{figure}

\subsection{Nuclear Matrix Elements}
Nuclear-structure calculations employed the NuShellX code with the jj45 model space and jj45pnb interaction. Energy levels of $^{113}$In were computed to match experimental states at 1024.280 keV and 1029.650 keV. The theoretical state at $E^{*}_{\rm th} = 1132$ keV ($J_f = 5/2$) best corresponds to the former. According to data the experimental state at 1029.650 keV can either be of spin $J=1/2^+$ or $J=3/2^+$. The best matches to these spin states are a computed $J=1/2^+$ state at 1843 keV and a computed $J=3/2^+$ state at 1891 keV of excitation. The  calculated half-lives corresponding to these states are displayed in Table~\ref{table:T12-allowed}. The experimental (partial) half-life of the decay transition to the 1029.650 keV state is 30613 yr \cite{NNDC} which is a rough two-to-three orders of magnitude longer than the computed ones but points to a spin-parity assignment $J=1/2^+$. The large discrepancy between the computed and experimental partial half-lives could be assigned to the inaccuracies of the adopted NSM Hamiltonian when applied to very weak EC transitions to highly excited states and the experimental uncertainties in determining the branching ratios of very weak EC transitions.

Table~\ref{table:T12-allowed} presents predicted half-lives for the EC decay of $^{113}$Sn to the excited states of $^{113}$In at 1024.280 keV and 1029.650 keV. Figures~\ref{fig:4-spectra} and~\ref{fig:5-zoom-spectra} show the normalized energy-release distributions. The transition $^{113}$Sn (1/2$^{+}$) $\rightarrow$ $^{113}$In$^*$ (1024.280 keV), with $Q_{EC}^* = 14.97$ keV, has $\Delta_{\rm L1} = 10.95$ keV. The transition to 1029.650 keV, with $Q_{EC}^* = 9.60$ keV, is closer to the L1 shell ionization energy ($\Delta_{\rm L1} = 5.58$ keV), showing stronger resonance enhancement near the endpoint, making it a preferred candidate for neutrino-mass studies.

Figure~\ref{fig:4-spectra} presents the normalized distribution of energy released after the decay of $^{113}$Sn to the excited states of $^{113}$In with the $Q_{EC}^* = 9.60$ keV for the allowed transition and for $^{163}$Ho to the ground state of $^{163}$Dy. In Fig.~\ref{fig:5-zoom-spectra} the enlarged datail of the interest region from Fig.~\ref{fig:4-spectra} is presented. As can be seen, the spectrum for $^{113}$Sn is three orders of magnitude lower than the spectrum for $^{163}$Ho; thus, it is not the most suitable for neutrino-mass determination.

\section{Conclusion and outlook}
A high-precision measurement of the gs-to-gs EC $Q$ value for $^{113}$Sn (1/2$^{+}$) $\rightarrow$ $^{113}$In (9/2$^{+}$) was performed using the PI-ICR technique at JYFLTRAP, yielding $Q_{EC}$ = 1039.25(19) keV, an eightfold precision improvement over AME2020. The $^{113}$Sn mass excess, $-$88327.87(27) keV/$c^2$, is six times more precise. Two gs-to-es EC transitions to excited states  at 1024.280 keV and 1029.650 keV in $^{113}$In were confirmed to be  energetically allowed. The 1029.650-keV transition, with $\Delta_{\rm L1} \approx 5.58$ keV, turns out to be promising for neutrino-mass experiments due to its allowed nature and a relatively short half-life. Further precision measurements of the branching ratio of the transition are needed. Theoretical predictions using the DHFS method and the nuclear shell model, including also exchange, overlap, shake-up, and shake-off corrections, support these findings.

\acknowledgments 
We acknowledge the staff of the Accelerator Laboratory of University of Jyv\"askyl\"a (JYFL-ACCLAB) for providing stable online beam. We thank the support by the Academy of Finland with  projects No. 314733,  315179, 327629, 320062, 354589, 345869 and 354968.
The support by the EU Horizon 2020 research and innovation program under grant No. 771036 (ERC CoG MAIDEN) is acknowledged.  
V.A.S., S.S., J.S., and J.K. acknowledge support from the NEPTUN project (PNRR-I8/C9-CF264, Contract No. 760100/23.05.2023 of the Romanian Ministry of Research, Innovation and Digitization). O. N. acknowledges support from the Institute of Atomic Physics through the CERN/ISOLDE grant, and from the Romanian National Authority for Research through the Nucleu Project No. PN 23 21 01 02.
This project has received funding from the European Union’s Horizon Europe Research and Innovation programme under Grant Agreement No 101057511 (EURO-LABS).

\bibliography{my-final-bib-from-jabref}

\end{document}